\documentclass{PoS}

\usepackage[T1]{fontenc}
\usepackage{newtxtext,newtxmath}
\usepackage{ae,aecompl}
\usepackage{graphicx}	% Including figure files
\usepackage{amsmath}	% Advanced maths commands
\usepackage{amssymb}	% Extra maths symbols
\usepackage{footmisc}
\usepackage{url}
\usepackage{subfig}
\usepackage{algpseudocode}
\usepackage[noend]{algorithm}
\usepackage{float}
\algtext*{EndIf}

\title{Analysis of differential observations of the cosmological radio background : studying the SZE-21cm}

\ShortTitle{Differential Observation Techniques for the SZE-21cm}

\author{\speaker{Charles Mpho Takalana}\thanks{A footnote may follow.}\\
        University of the Witwatersrand\\
        E-mail: \email{mtakalana@ska.ac.za}}

\author{Sergio Colafrancesco\\
        University of the Witwatersrand\\
        E-mail: \email{Sergio.Colafrancesco@wits.ac.za}}

\author{Paolo Marchegiani\\
        University of the Witwatersrand\\
        E-mail: \email{Paolo.Marchegiani@wits.ac.za}}

\abstract{In pursuit of understanding the early Universe and early processes occurring particularly in the Dark Ages (DA) and the Epoch of Reionization (EoR), it is vital that a suitable probe is identified. Probing these epochs will be useful in studies of the origin of first galaxies and most importantly formation of early black-holes. Over the past decade numerous probes have been proposed, with one of the most promising being the SZE-21cm, a specific form of the Sunyaev-Zeldovich effect (SZE) produced when photons of the 21cm background are inverse Compton up-scattered by electrons residing in hot plasma of cosmic structures such as galaxy clusters and active radio galaxies. The SZE-21cm is calculated in a full relativistic approach  of the scattering processes of the CMB photons modified by the cosmological redshifted 21cm background in the hot intra-cluster medium of galaxy clusters capturing effects induced by relativistic corrections to this scattering and by multiple scattering effects. We apply image differencing techniques to simulated radio observations of galaxy clusters using the redshifted 21cm background, we conduct this making use of the public semi-numeric code 21cmFAST.  We are able to achieve subtraction of contaminating foregrounds through pixel by pixel operations on the data retrieved from our simulated data cubes. We demonstrate that SZE-21cm can be recovered through differential observations of the 21cm background.}

\FullConference{5th Annual Conference on High Energy Astrophysics in Southern Africa\\
		4-6 October, 2017\\
		University of the Witwatersrand (Wits), South Africa}

\begin{document}

\section{Introduction}
\noindent
Studies of the Dark Ages (DA) and the Epoch of Reionization (EoR) of the Universe will shed light on a large number of fundamental questions in cosmology that will help us understand the properties of the first galaxies and physical mechanisms that give rise to stuctures we observe today. The generally accepted probe for these epochs is the 21cm spectral line produced by a spin flip in neutral hydrogen \cite{van,Muller,ewen}. Low frequency observations of the redshifted 21cm line of neutral hydrogen present promising avenues for exploring periods in the Universe that are to date poorly constrained. We have yet to observe the period of formation of the first luminous structures (DA), 20 $\lesssim$ $z$ $\lesssim$ 1100, and have recently started making ground in exploring the era of the first light that stretches from the formation of these structures to the complete reionization of the intergalactic medium (IGM), 6 $\lesssim$ $z$ $\lesssim$ 20. The detectable signal in the frequency range relevant to the DA and EoR is composed of a number of components, namely: instrument response, instrument noise, ionospheric distortions, galactic and extragalactic foregrounds as well as the cosmological signal itself, which is expected to vary with redshift \cite{Ciardi,Mellema}. Foreground emission is expected to be 4-5 orders of magnitude stronger than the cosmological signal \cite{Shaver}, unscrambling the signal and foregrounds will be non-trivial and foreground subtractions may leave contaminating residuals. Foregrounds will have to be carefully removed with the use of highly accurate and precise cleaning methods as any error could potentially destroy the 21cm cosmological signal. 
\\
The Sunyaev-Zeldovich Effect (SZE) \cite{SZ, SZ1} is a powerful cosmological tool that has been used to identify galaxy clusters; this signal does not vary with distance from the source due to its redshift independent nature. Combining the advantages of the 21cm line and the SZE creates a potential probe suitable for studying the Universe at the EoR and DA with the highest precision, we refer to this probe as the SZE-21cm. Coorey 2006 \cite{Cooray} and Colafrancesco et al. 2016 \cite{Colafrancesco} emphasised that differential observations of the SZE-21cm with radio interferometers are less affected by the presence of galactic and extra-galactic foregrounds that are uniform over angular scales larger than a typical cluster and the observations would be less affected by exact calibration of the observed intensity using an external source. Proposed observations of the SZE-21cm can be carried out with interferometers pursued for studying and imaging the 21cm line of neutral hydrogen through differential observations toward and away from galaxy clusters (e.g. Low Frequency Array (LOFAR)\footnote{\url{http://lofar.org}}, Murchison Widefield Array (MWA)\footnote{\url{http://www.mwatelescope.org}}, Hydrogen Epoch of Reionization Array (HERA)\footnote{\url{http://reionization.org }}, Square Kilometre Array (SKA)\footnote{\url{http://www.skatelescope.org}}). Once complete the SKA precusor, HERA, with its 350 elements opperating at frequencies from 50 to 250 MHz will be the first instrument with potential to directly image the EoR \cite{DeBoer}. In future SKA-1 Low will cover the 50-350 MHz frequency range, consisting of nearly 130 000 antennas across 500 stations with a maximum 65 km baseline \cite{Dewdney}, allowing us to image the EoR and DA. The technique we study here for the SZE-21cm will be applicable to images obtained from both HERA and SKA-1 Low.
\\
This work builds on Colafrancesco et al. 2016 \cite{Colafrancesco} where the theoretical aspect was laid out. We apply the theory here to image differencing of simulated observations of the 21cm cosmological signal made using 21cmFAST\footnote{\url{homepage.sns.it/mesinger/}}. We attempt to recover the SZE-21cm signal from simulated data, and use this to construct expectations for low frequency differential observations. This work links theory, imaging technique and analysis. 
\\
This paper is structured as follows. Section \ref{21cmSZE-21cm} unpacks the SZE-21cm signal and formalism for differential observations of the SZE-21cm as well as the 21cm cosmological signal itself. Section \ref{SZE-21cm} goes through the methods and results of mock observations of the SZE-21cm and section \ref{Discussion} presents the discussion. We conclude in section \ref{conclusion}. Throughout this paper we assume {$\Lambda$}CDM-cosmology with parameters: $h$ = 0.673, $\Omega_{m}$ = 0.315 ,  $\Omega_{b}$ = 0.0491, $\Omega_{\Lambda}$ = 0.685, $\sigma_{8}$ = 0.815 and $n_{s}$ =  0.968.

\section{Probing the DA and EoR with the SZE-21cm}
\label{21cmSZE-21cm}

\subsection{The 21cm Cosmological Signal}
\noindent
The 21cm line is a result of the ground state hyperfine transition of atomic hydrogen, emitting radio waves at 21cm, a rest frequency $\nu_{0}$ = 1420.4057 MHz. The 21cm line allows for mapping of large scale structures including Magellanic clouds, galaxies and galaxy clusters. The 21cm signal is expected to be observed as a faint, diffuse background at low frequencies ($\lesssim$ 200 MHz) \cite{Colafrancesco}, its detection will yield information about the global and local properties of the Inter Galactic Medium (IGM). The spin temperature, $T_{s}$, quantifies the excitation temperature for the 21 cm transition, with $n_{1}/n_{0}$ = 3 exp (- $T_{*}$/$T_{s}$) representing the ratio of the electron density in the triplet state ($n_{1}$) to the singlet state ($n_{0}$) of the hyperfine level. $T_{*}$ is the temperature corresponding to the 21cm wavelength ({$T_{*}$}= $\frac{h\nu_{0}}{k_{B}}$ = 0.0681 K), and ${g_{1}}/{g_{0}}$ = 3  is the ratio of the spin degeneracy factors of the two levels. The spin temperature essentially is a shorthand for the ratio between the occupation number of the two hyperfine levels which establishes the intensity of the radiation emerging from a cloud of neutral hydrogen \cite{Zaroubi} and is determined by three processes: absorption of CMB photons; collisions between the particles, which is defined by the gas kinetic temperature, $T_{k}$; and scattering of ambient UV (Ly-$\alpha$) photons, $T_{\alpha}$. Assuming $T_{\alpha}$ = $T_{k}$ (Field et al. 1958 \cite{Field}):

\begin{align}
\hspace*{5mm} T_{S} = \frac{T_{CMB} + y_{k}T_{k} + y_{\alpha}T_{k}}{1 + y_{\alpha} + y_{k}}
\label{eq:eq12}
\end{align}

\noindent
$y_{\alpha}$ and $y_{k}$ are the Ly-$\alpha$ and kinetic coupling terms respectively. The Ly-$\alpha$ coupling term $y_{\alpha}$  is due to the Wouthuysen-Field process (Ly{$\alpha$} pumping) occurring through collisions and scattering of Ly{$\alpha$} photons coupling the spin temperature to the gas temperature, {$T_{k}$}, making neutral hydrogen visible in absorption or emission as a result of the gas being colder or hotter than that of the CMB \cite{Evoli, Field, Hirata, Wouthuysen}. Changes in the CMB temperature, the kinetic temperature, and spin temperature (${T_{s}}$) lead to interesting physical features of the HI 21cm signal. The quantity that 21cm interferometers will measure is the 21cm differential brightness temperature $\delta T_{b}$ $\equiv$ $T_{b}$ - $T_{CMB}$, a deviation from the CMB temperature also known as the global brightness temperature \cite{Zaroubi}

\begin{align}
\hspace*{5mm} \delta T_b \approx 25x_{HI}(1+\delta) \left(\frac{1+z}{10} \right) ^{1/2} \left[1 - \frac{T_{CMB}}{T_{s}}\right] \left[\frac{H(z)/(1+z)}{d\nu_\| / dr_\|}\right] mK.
\label{eq:eq10}
\end{align}

\begin{figure}
	\centering
	\subfloat[]{\includegraphics[width=3.1in]{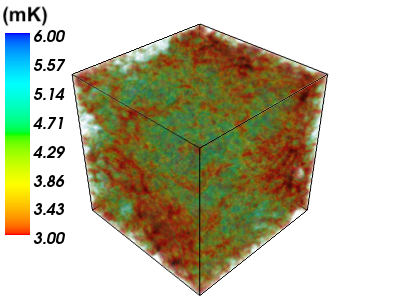}}
	\subfloat[]{\includegraphics[width=3.1in]{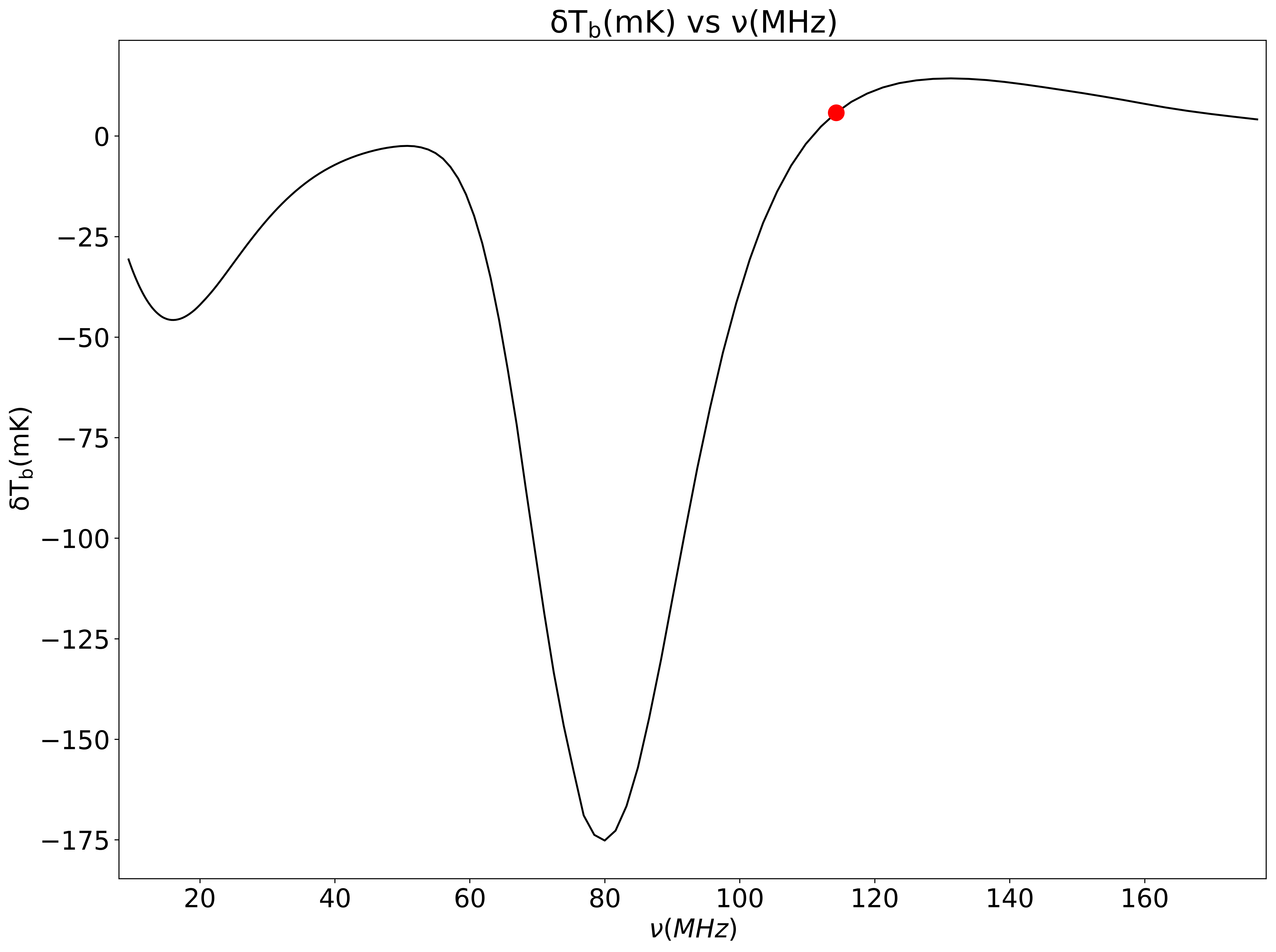}} 
	\caption{\textit{\textbf(a) Simulated 21cm Brightness temperature 3D volume at 114.27 MHz.  \textbf(b) The sky averaged 21cm brightness temperature offset from the CMB spectrum as a function of frequency from our simulations.}}
	\label{fig:21cm} 
\end{figure}

\noindent
Equation \ref{eq:eq10} is a mixture of cosmology and astrophysics dependent terms. $\delta T_{b}$ is controlled by different contributions at different stages of its evolution. At high redshifts (z $\gtrsim$ 10) the neutral fraction ($x_{HI}$) is approximately one, $\delta T_{b}$ is proportional to the density fluctuations, and this measurements places it as an excellent cosmological probe. At low redshifts (z $\lesssim$ 10) the Universe is ionized and  measurements of $\delta T_{b}$ are dominated by the contrast between the neutral and ionized regions, and this probes the astrophysical source of ionization. Equation \ref{eq:eq10} tells us that the 21cm radiation can only be observed if it has a temperature that differs from that of the CMB, no signal is expected  when $T_{S}$ $\sim$ $T_{CMB}$ \cite{Field, Furlanetto, Wouthuysen}. 
\\
\noindent
Semi-numeric modeling tools such as 21cmFAST \cite{Mesinger} and Simfast21 \cite{Santos1, Santos2} simulate the observable 21cm differential brightness temperature $\delta T_{b}$ which is detected differentially as a deviation from the CMB temperature. The 21cm global brightness temperature signal we expect to observe is in figure \ref{fig:21cm}, this was obtained from our simulation run of 21cmFAST which we use as the basis of this text, an example of a simulated 30 Mpc data cube at 114.27 MHz is presented along side the spectrum, figure \ref{fig:21cm}.

\subsection{The SZE-21cm}
\noindent
Inverse Compton scattering slightly alters the incident Planck spectrum, changing the apparent brightness of the Cosmic Microwave Background (CMB) radiation towards a reservoir of hot plasma \cite{Birkinshaw1999} imprinting on it a unique spectral signature known as the Sunyaev Zeldovich effect (SZE) \cite{SZ, SZ1, SZ2, SZ3}. Initially proposed by Cooray 2006 \cite{Cooray} and later redefined taking a relativistic approach by Colafrancesco et al. 2016 \cite{Colafrancesco}, the SZE-21cm is an alternative method that can be used to overcome foreground challenges faced when studying the early epochs using the 21cm line because it is a differential measure of the radio background spectrum on and off an area of the sky containing the cosmic structure of interest. The SZE-21cm is the spectral distortion of the CMB spectrum altered by physical effects occurring during the epoch related to the emergence of the 21cm radiation background, photons of the CMB radiation are scattered by electrons in the hot intracluster gas.  Differential observations of this signal with low frequency radio interferometers are less affected by large-scale foregrounds and exact calibration of the observed intensity \cite{Colafrancesco, Cooray}. 
\\
In this section we go through the derivation for the SZE-21cm, the reader is refered to Colafrancesco et al. 2003 and 2016 \cite{Colafrancesco, Colafrancesco1} for full derivations in the relativistic limit. The scattering optical depth of a cluster which contains gas integrated along the line of sight of the electron concentration $n{_e}($\textit{\textbf{r}}$)$ is given by

\begin{align}
\hspace*{5mm}\tau{_e} = \int{n{_e}(\textit{\textbf{r}}) \sigma{_T} d\textit{l}} 
\label{eq:eq1}
\end{align}

\noindent
where $\sigma_T$ is the Thomson cross section; this is required to calculate the electron redistribution function \textit{P(s)} as the Poissonian probability containing relativistic corrections essential for correctly describing the Compton scattering produced by relativistic electrons in hot plasma

\begin{align}
\hspace*{5mm} P(s) = \sum_{n=0}^{+ \infty} \frac{e^{-\tau}\tau^{n}}{n!}P_{n}(s) 
\label{eq:eq2}
\end{align}

\noindent
where $P_{n}(s)$  is the probability function to have $n$ scatterings that is given by the convolution of $n$ times the single scattering probability $P_{1}(s)$ (e.g. Colafrancesco et al. 2003 \cite{Colafrancesco1}), and the inverse Compton scattering process yields the probability of a logarithmic shift $s= ln(\frac{\nu'}{\nu})$ in the photon frequency. The general form of the  spectral distortion of the CMB due to the SZE is given by

\begin{align}
\hspace*{5mm}I(x) = \int_{-\infty}^{+\infty} I_{0}(xe^{-s}) P(s) ds
\label{eq:eq3}
\end{align}

\noindent
where the normalised frequency is represented as ${x = \frac{h\nu}{kT_{CMB}}}$ and $I_{0}$ is the specific intensity of the incident CMB radiation field \cite{Colafrancesco, Colafrancesco1}. The general form of the SZE is than given by the difference:

\begin{align}
\hspace*{5mm}\Delta I(x) = I(x) - I_{0}(x).
\label{eq:eq4}
\end{align}

\noindent
For the standard case the incoming spectrum is the standard CMB spectrum:

\begin{align}
\hspace*{5mm}I_{0,st}(x) = 2 \frac{(KT_{0})^{3}}{(hc)^{2}} \frac{x^{3}}{e^{x}-1}:
\label{eq:eq5}
\end{align}

\noindent
putting this into equations \ref{eq:eq3} and \ref{eq:eq4} we get the standard SZE, ${\Delta}{I}_{st}(x)$.
\noindent
Calculations that take into consideration modifications to the CMB spectrum by neutral hydrogen can be made to establish the low frequency spectrum emerging from a galaxy cluster post scattering. The expected modification to the input spectrum is a few milli-Kelvin correction to the temperature of the CMB black body spectrum.  Coorey 2006 \cite{Cooray} gave a description of the SZE-21cm in a non-relativistic approximation neglecting effects induced by relativistic corrections, these are instead accounted for by Colafrancesco et al. 2016 \cite{Colafrancesco}. The CMB spectrum modified by neutral hydrogen during the DA and EoR is given by

\begin{align}
\hspace*{5mm}I_{0,mod}(\nu) = I_{0,st}(\nu) + \delta I(\nu) 
\label{eq:eq6}
\end{align}

\noindent
where ${\delta}I(\nu)$ can be written in terms of the brightness temperature change relative to the CMB

\begin{align}
\hspace*{5mm}{\delta}T_{b}(\nu) = \frac{c^{2}}{2k\nu^{2}} \delta I (\nu).
\label{eq:eq7}
\end{align}

\noindent
With $I_{0,mod}(\nu)$ as our incoming radiation spectrum and the SZE-21cm reads

\begin{align}
\hspace*{5mm}\Delta I_{mod}(\nu) = I_{mod}(\nu) - I_{0,mod}(\nu).
\label{eq:eq8}
\end{align}

\noindent
Applying equation \ref{eq:eq7}, we can rewrite equation \ref{eq:eq8} in terms of temperature as

\begin{align}
\hspace*{5mm}\Delta T_{mod}(\nu) = T_{mod}(\nu) - T_{0,mod}(\nu).
\label{eq:eq9}
\end{align}

\section{Differential Analysis techniques for the SZE-21cm}
\label{SZE-21cm}
\noindent
This work makes use of 21cmFAST \cite{Mesinger} which employs semi-numerical approaches to produce large-scale simulations making use of: analytical prescriptions, the excursion set formalism (used to predict the clustering properties of dark matter halos relative to the dark matter), and perturbation theory to produce evolved 3D realizations of spin temperature fields, density, peculiar velocity and ionization, which when combined can be used in the computation of the 21cm brightness temperature described by equation \ref{eq:eq10}. These simulations produce ${\delta}T_{b}$ cubes (signal as a function of frequency). We produced three dimensional 30 $h^{-1}$ comoving Mpc simulation cubes, from redshift z = 6.89 up to redshift z = 149.74, these all make up our mock observations. The initial grid of the cubes is $460^{3}$ voxels which are smoothed down to a $115^{3}$ grid. The standard CMB required for analysis of the SZE is not taken into account by 21cmFAST and therefore we add the standard CMB to the 21cm brightness temperature in all the simulated cubes through equation \ref{eq:eq6}; the result of this are CMB cubes modified by neutral hydrogen during the DA and EoR.

\subsection{Temperature variations in 21cm simulated observations}
\begin{figure}
	\centering
	\subfloat[]{\includegraphics[width=2.6in]{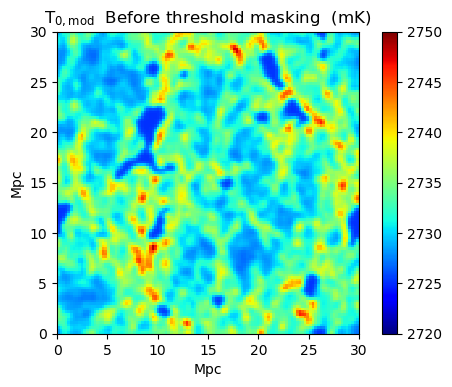}}
	\subfloat[]{\includegraphics[width=2.7in]{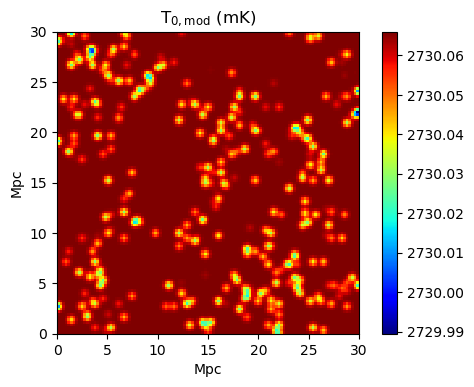}} 
	\caption{\textit{\textbf{a}: Original 21cm background image from an area in the simulated cube before pixel replacement with temperature variations of order  $10^{-1}$ -  $10^{-3}$ at 114.27 MHz, \textbf{b}: The 21cm background image after pixels were replaced such that the cube variations were of order $10^{-5}$ at 114.27 MHz.}}
	\label{fig:variation} 
\end{figure}

\noindent
Temperature variations in the ${\delta}T_{b}$ cubes were of order  $10^{-1}$ -  $10^{-3}$, the order of these variations create a great concern as they lead to incorrect values for the SZE-21cm when conducting our differential observations. This arises from the SZE-21cm being a CMB effect and that the CMB primodial fluctuations are of order $10^{-5}$, hence the background we substract through the differencing procedure need to contain similar order variations to the CMB. The source of variations in these cubes are significant degeneracies that  exist between parameters that govern the mean brightness spectrum leading to inhomogeneities which can be observed by interferometers pursued for 21cm observations \cite{McQuinn}. The inhomogeneities are induced by fluctuations from various regions of the IGM which include contributions from different physical properties including velocity gradients, temperature (gas \& spin), ionization state and density. Establishing signatures associated with the mean brightness spectrum independent of spacial variations could play a role in breaking some of the degeneracies. Inhomogeneities generate anisotropies in the brightness temperature during the DA and EoR \cite{McQuinn}, in addition to this there are fluctuations due to gravitational lensing through scattering in galaxies. We do not however focus on how the degeneracies can be broken as this is beyond the scope of this text. We employ pixel replacements which not only take care of variations due to inhomogeneity but also bright sources and foregrounds. In the context of this work bright sources in the background image were identified as foreground contaminants, and hence were removed from the maps to prevent them from contributing to the brightness temperature fluctuations. 
\noindent
Taking a cube slice away from the inserted cluster as the background image: pixels $T_{0,mod_{i}}$ in the background image were replaced with the mean cube temperature $\overline{T}_{0,mod}$ as described below: 
\\
\begin{algorithmic}
	\centering
	\If {$\frac{T_{0,mod_{i}} - T_{cmb}}{T_{cmb}} \gtrsim 10^{-5}$}
	\State $T_{0,mod_{i}} = \overline{T}_{0,mod}$
	\EndIf
\end{algorithmic}
\noindent
this was done to ensure that the background remains with variations of the same order as the CMB. Figure \ref{fig:variation} shows a background image before and after the variations are dealt with.

\subsection{Method for simulated differential observations for the SZE-21cm}
\begin{figure}
	\centering
	\subfloat[]{\includegraphics[width=2.7in]{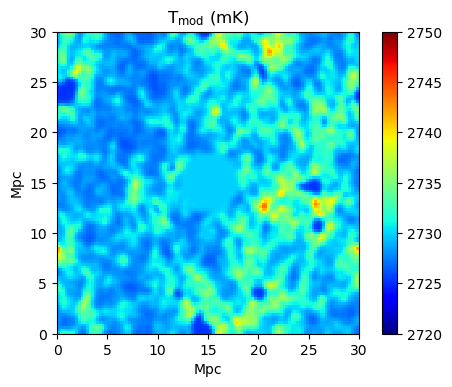}}
	\subfloat[]{\includegraphics[width=2.7in]{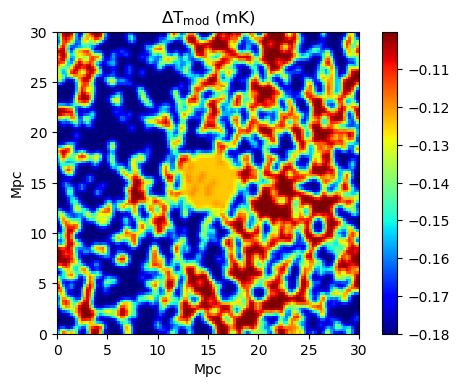}}
	\caption{\textit{In (\textbf{a}) we have an image of the inserted cluster in the presence of a 21cm background at 114.27, and in (\textbf{b}) we have the final image of the cluster SZE-21cm obtained from the image differencing procedure at 114.27 MHz.}}
	\label{fig:cluster} 
\end{figure}

\noindent
Our differential observing technique hinges on equation ~(\ref{eq:eq9}), $T_{mod}(\nu)$  represents the patch of sky that contains a galaxy cluster, and $T_{0,mod}(\nu)$ represents a patch of sky containing only the background. The cluster signal we insert into our cube was calculated using the mean 21cm brightness temperature of each cube through the spectrum in figure \ref{fig:21cm}, this cluster signal is $T_{mod}(\nu)$ in equation \ref{eq:eq9}. The inserted spherical cluster in figure \ref{fig:cluster} (a) has a temperature of 10 keV, optical depth of {$10^{-3}$} and a radius of 3 Mpc. These values are typical of rich clusters where the SZE has been observed; the SZE we consider at the cluster is due only to the thermal effect. The slice away from the cluster represents our  background in figure \ref{fig:variation} (b) with the temperature variations taken care off. Following the subtraction of the background image from the cluster image through equation \ref{eq:eq9} we are left with the SZE-21cm signal at the cluster and residual foregrounds, the result image of this subtraction is figure \ref{fig:cluster}(b) .

\noindent
The differencing procedure was followed for cubes of frequency range 6 to 180 MHz, the SZE-21cm was measured in each cube and used to construct the spectrum in figure \ref{fig:sze21cm}  which we discuss in section \ref{Discussion}.

\section{Discussion on  SZE-21cm simulation results}
\label{Discussion}
\begin{figure}
	\centering
	\includegraphics[width=4in]{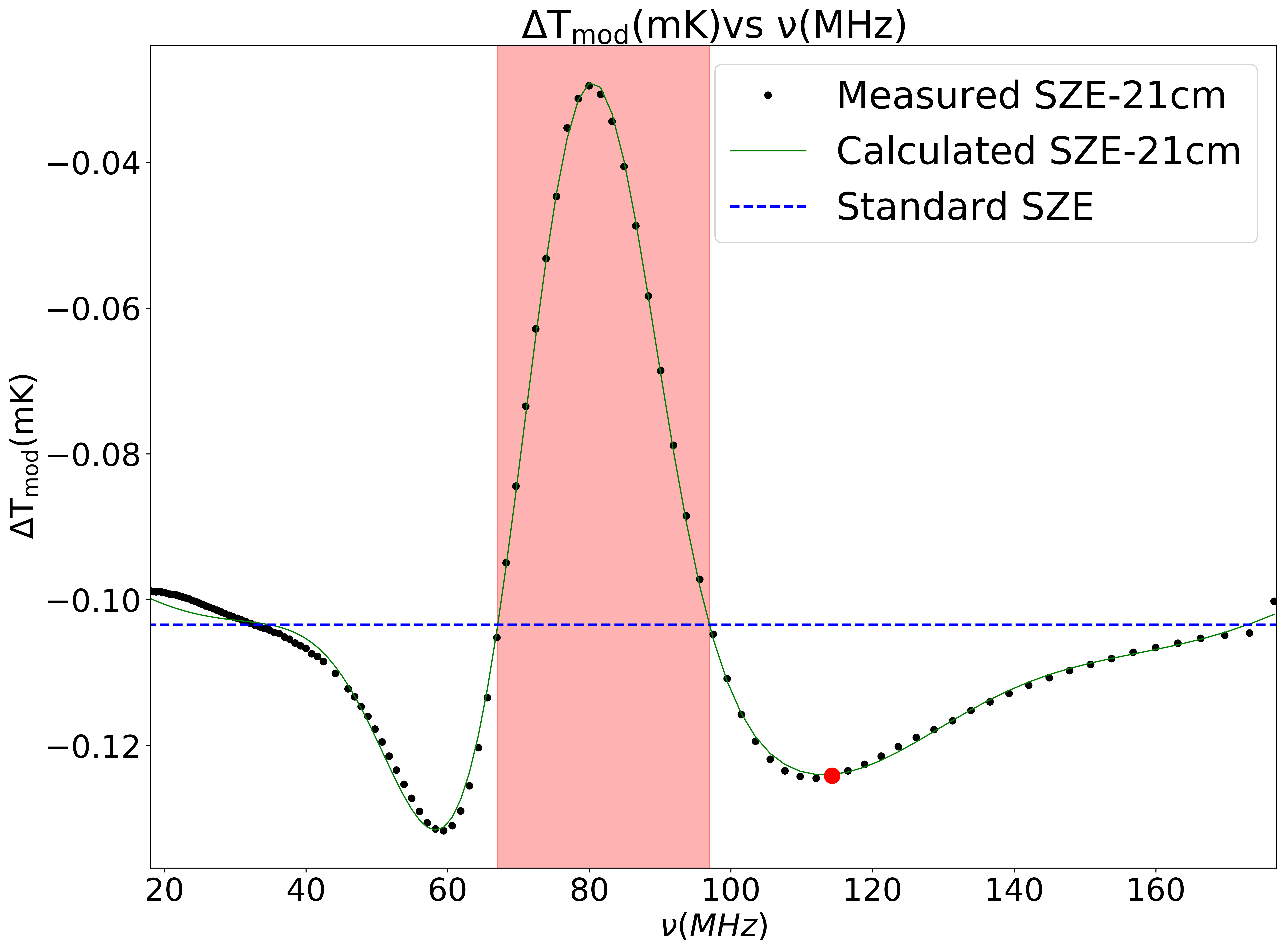}
	\caption{\textit{The SZE-21cm measured following the image differencing technique for a cluster of 10 keV, an optical depth of {$10^{-3}$}. The red area represents where the SZE-21cm is bigger than the standard SZE. The green line is the best-fit line of theoretically calculated data points (Colafrancesco et al. 2016 \cite{Colafrancesco}), the black dots are points measured from the images and the blue dashed line is the standard SZE. The red point corresponds to the sample figure at 114.27 MHz.}}
	\label{fig:sze21cm} 
\end{figure}

\noindent
The SZE-21cm can be used to study in depth the physical history of the EoR and DA. The SZE-21cm was obtained by measuring the signal after image differencing in section \ref{SZE-21cm}, we now discuss how this signal can be used to directly establish the global 21cm spectrum, and its dependencies on the properties of the plasma in galaxy clusters. Figure \ref{fig:sze21cm}  shows us the SZE-21cm spectrum obtained by following the differencing procedure in the previous section from 6 to 180 MHz. 

\noindent
The observables in the spectrum are listed below:

\begin{itemize}
	\item The SZE-21cm measured at the cluster (black dots in figure \ref{fig:sze21cm}) agrees closely with the calculations done in Colafrancesco et al. 2016 \cite{Colafrancesco}, represented by the green line. This is expected as the formalism used in the calculations is the same as that used for the simulations.
	
	\item The SZE-21cm is bigger than the standard SZE at 67 $\lesssim$ $\nu$ $\lesssim$ 97 MHz. The standard SZE is bigger than the SZE-21cm at $\nu$ $\lesssim$ 67 and 97 $\lesssim$ $\nu$ $\lesssim$ 180.
\end{itemize}
\noindent
Something that immediately stands out about the SZE-21cm spectrum is its resemblance of an inverted 21cm brightness temperature input spectrum, figure \ref{fig:21cm}. It is in this feature that we are able to use the SZE-21cm spectrum to narrate evolution of the 21cm signal itself. At frequencies where the curvature of the input 21cm brightness temperature spectrum is negative, a smaller number of photons are present at neighbor frequencies  with respect to the standard CMB spectrum. For frequencies 30 $\lesssim$ $\nu$ $\lesssim$ 67 MHz and 97 $\lesssim$ $\nu$ $\lesssim$ 174 MHz, the input spectrum has negative curvature, here the SZE-21cm is smaller than the standard SZE, where else for frequencies 67 $\lesssim$ $\nu$ $\lesssim$ 97 MHz the input spectrum has positive curvature and the SZE-21cm is bigger than the standard SZE. 
\noindent
Through the subtraction procedure guided by equation \ref{eq:eq9} we find that the minimum point of the input spectrum, figure \ref{fig:21cm}, corresponds to the maximum point of the SZE-21cm spectrum, figure \ref{fig:sze21cm}. Where a smaller number of photons exist with respect to the standard CMB, the resulting emission of the SZE-21cm is bigger than the standard SZE. Where the input spectrum is maximum, the SZE-21cm spectrum is at its minimum, a larger quantity of photons exist with respect to the standard SZE. The result difference images show some features associated with the subtracted background, these features are dominantly seen inside and outside the cluster region where the standard SZE is bigger.
	
\subsection{Detectability of the signal with SKA telescopes}
\noindent
The SZE-21cm and standard SZE are detectable with the SKA instrument; figure 5(a)  shows detectability of the non-modified SZE spectrum, $\Delta{I_{st}}$, and modified SZE spectrum, $\Delta{I_{mod}}$, compared with the sensitivities of SKA telescopes, SKA-50\%, SKA1-Low, and SKA-2. We use the SKA1 System Baseline document to extract the instrument performances \cite{Dewdney}. Observing at 100 kHz bandwidth, with 1000 hours integration time, 2 polarizations, no taper, no weight: the SZE-21cm for a 10 keV cluster will be detectable with SKA-50\% at frequencies $\gtrsim$ 90 MHz, SKA1-Low at frequencies $\gtrsim$ 85 MHz, and SKA2 will be able to detect the signal below 60 MHz in the DA. Figure 5(b) gives us an idea of the frequency bands where we will be able to distinguish between the SZE-21cm and the standard effect, that is the detectable difference. The SZE-21cm is weaker compared to the standard SZE due to the Wouthuysen \cite{Wouthuysen} effect at around 80 MHz, this is a good frequency to detect the difference between the two signals; another range is  at 110 - 120 MHz, where the SZE-21cm appears stronger due to ionization effects. With SKA1-low sensitivity at its high frequency band, the best frequency range to study the SZE-21cm is $\gtrsim$ 100 MHz. With SKA2 however we expect improved results, the difference between the SZE-21cm and the standard effect will be detectable at $\nu$ $\gtrsim$ 75 MHz.

\begin{figure}
	\subfloat[]{\includegraphics[width=2.8in]{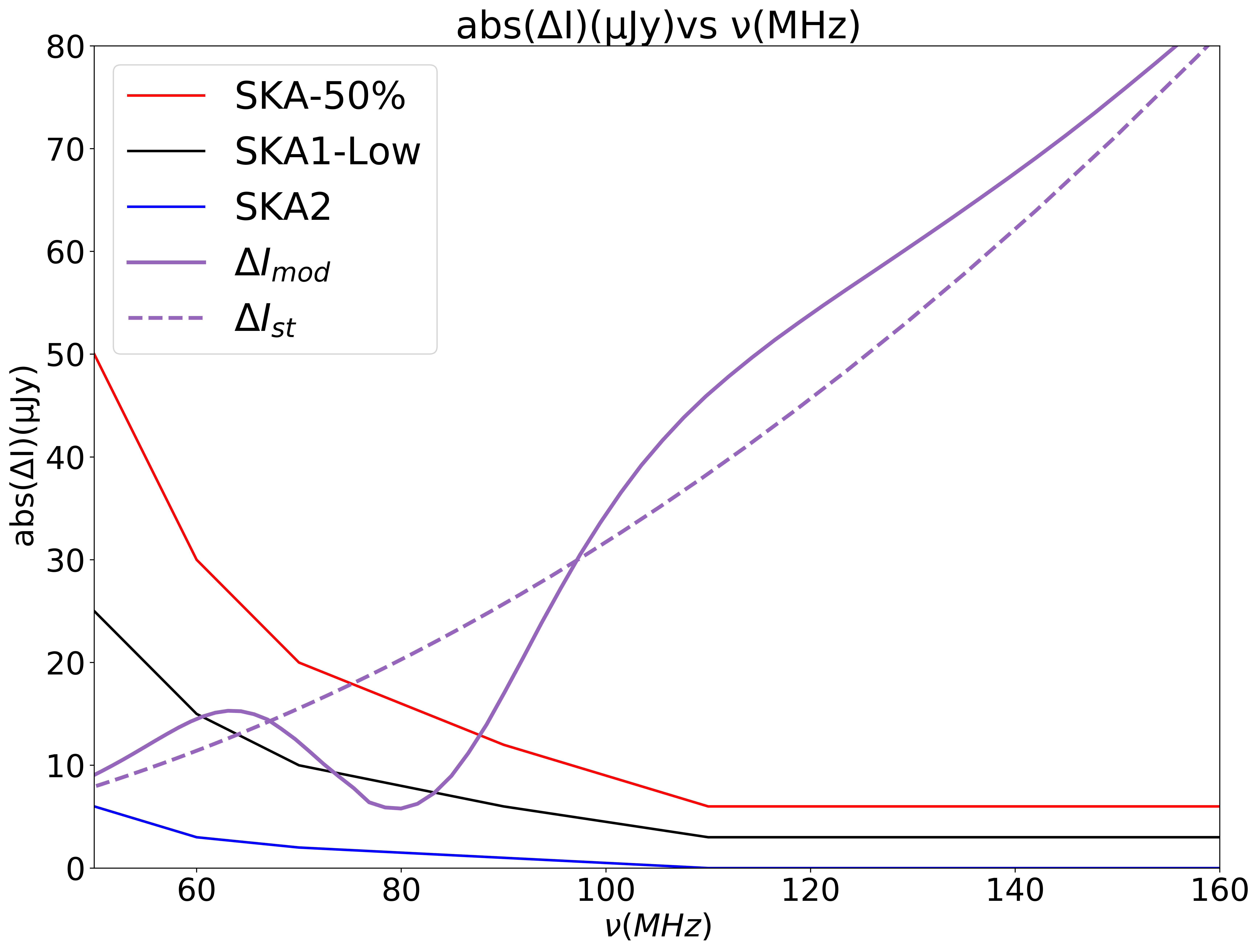}}
	\subfloat[]{\includegraphics[width=2.9in]{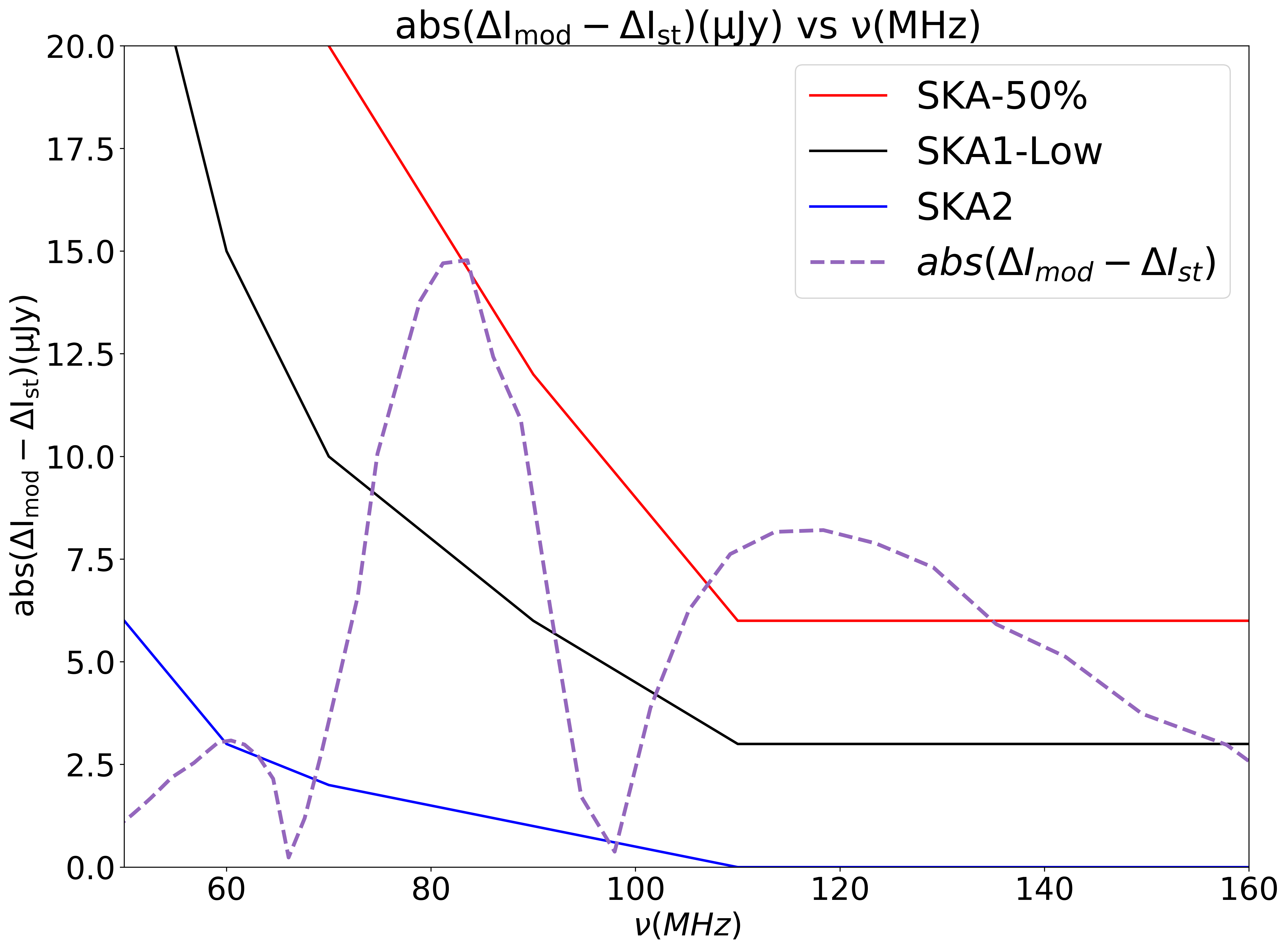}} 
	\caption{\textit{(\textbf{a}) Sensitivity of SKA telescopes to the SZE-21cm and the standard SZE, the dashed purple line is the Standard SZE, the solid purple line is the SZE-21cm. (\textbf{b}) Sensitivity of SKA telescopes to difference of the SZE-21cm and standard SZE, the dashed purple line is the absolute difference between the SZE-21cm and the standard SZE. In both figures the red, black and blue lines represent the telescope sensitivity as shown in the figure legend.}}
	\label{fig:EcUND} 
\end{figure}

\section{Conclusion}
\label{conclusion}
\noindent
The application of the differential technique to simulated 21cm observations shows that the SZE-21cm can be detected and measured directly from observation and may therefore be used to construct the global 21cm cosmological signal. The technique can further be bettered by averaging over a larger number of clusters which will be useful in dealing with the statistics of the SZE-21cm. 

\section{Acknowledgements}
\noindent
This work is based on the research supported by the South African Research Chairs Initiative of the Department of Science and Technology and National Research Foundation of South Africa (Grant No 77948). C.M., S.C. $\&$ P.M. acknowledge support from the Department of Science and Technology/National Research Foundation (DST/NRF) Square Kilometre Array (SKA) post-graduate bursary initiative under the same Grant.

\end{document}